\documentclass[twocolumn,aps,prl,10pt,nofootinbib]{revtex4-1}%,nofootinbib]{revtex4-1}
\usepackage{amsmath}
\usepackage{amssymb}
\DeclareMathAlphabet{\mathsuet} {T1} {wesu}{bx}{sl}
\usepackage{mathtools}
\usepackage{graphicx}

\usepackage{hyperref}
\usepackage{dsfont}
\usepackage{newtxtext}
\usepackage[cmbraces]{newtxmath}
\DeclareMathSymbol{g}{\mathord}{lettersA}{188}
\DeclareMathSymbol{y}{\mathord}{lettersA}{189}
\DeclareMathSymbol{v}{\mathord}{lettersA}{190}
\DeclareMathSymbol{w}{\mathord}{lettersA}{191}
\usepackage{slashed}

\hypersetup{
    colorlinks,
    linkcolor={blue},
    citecolor={blue},
    urlcolor={blue}
}

\allowdisplaybreaks

\graphicspath{{./}{./figs/}{../figs/}{../Figs_trial/}{../Yukawa/}}

\usepackage[dvipsnames]{xcolor}
\usepackage[normalem]{ulem}

\begin{document}

\title{%
Non-Lagrangian phases of matter from Wilsonian renormalization of 3D Wess-Zumino-Witten theory on Stiefel manifolds
}

\author{Shouryya Ray}
\email{shouryyar@setur.fo}
\affiliation{Department of Science and Technology, University of the Faroe Islands, Vestara Bryggja 15, 100 T\'orshavn, Faroe Islands}

%%%%%%%%%%%%%%%%%%%%%%%%%%%%%%%%%%%%%%%%%%%%%%%%%%%%%%%%%%%%%%%%%%%%%%%
\begin{abstract}
I study the renormalization of $D$-dimensional level-$k$ Wess-Zumino-Witten theory with Stiefel-manifold target space $\operatorname{St}_{N,N-D-1} \cong \operatorname{SO}(N)/\operatorname{SO}(D+1)$, with a particular focus on $D=3$. I investigate in particular whether such a theory admits IR-stable fixed points of the renormalization group flow. Such fixed points have been suggested to describe conformal phases of matter that do not have a known dual (super-)renormalizable Lagrangian for $N \geqslant 7$ in $D=3$. They are hence of interest both from the point of view of quantum phases of matter as well as pure field theory. The $D$-dimensional expressions enable the computation, by analytic computation, of beta functions in $D=2+\epsilon$, at least to first non-trivial order. In $D=2$, a stable fixed point is found, serving a generalization of the famed $\operatorname{SU}(2)_k$ Wess-Zumino-Witten conformal field theory; it annihilates in $D=2+\epsilon$ with an unstable fixed point which splits off from the Gaussian one for $\epsilon > 0$. Although the story is thus qualitatively similar to that of SO(5) deconfined (pseudo-)criticality, for $N \geqslant 6$, the annihilation appears to occur only for $\epsilon > 1$, suggesting the existence of a stable phase in $D=3$. Comparisons of the scaling dimension of the lowest singlet operator are made with known results for $N=6$, which is dual to QED$_3$ with $N_\text{f}=4$ fermion flavors. The predictions for the $N=7$ Stiefel liquid represent to my knowledge the first computation of this kind for a Wess-Zumino-Witten theory without a known gauge theory dual.
\end{abstract}
%%%%%%%%%%%%%%%%%%%%%%%%%%%%%%%%%%%%%%%%%%%%%%%%%%%%%%%%%%%%%%%%%%%%%%%

\date{\today}

\maketitle

%%%%%%%%%%%%%%%%%%%%%%%%%%%%%%%%%%%%
%INTRODUCTION
%%%%%%%%%%%%%%%%%%%%%%%%%%%%%%%%%%%%
\paragraph{Introduction.} Duality is arguably one of the most intriguing features of modern quantum field theory. One presentation thereof, more precisely called IR duality, is that seemingly distinct renormalizable field theories can be described in the IR by the same effective, possibly non-renormalizable, field theory. From a complementary point of view, one often expects a non-renormalizable effective field theory, if realized in nature, to possess a renormalizable UV completion. One often intuitively pictures the degrees of freedom of the UV completion to be `more fundamental', at least compared to the degrees of freedom of the non-renormalizable effective field theory---a meson, for instance, is in reality a quark-antiquark pair held together by gluons. Duality is the statement that this ``factorization'' of effective degrees of freedom into more fundamental ones is not always unique. In the language of the renormalization group (RG), different UV theories can flow to the same IR theory: the renormalization group is in fact only a semigroup.

This point of view is especially pertinent in condensed matter physics. From experience, it is known that exciting novel phases can arise when there is an intertwinement of two (or more) orders \cite{Senthil:2005jk,FradkinColloquium}, i.e., when the topological defects of one phase are charged under the symmetry broken spontaneously in the other. (In other words, there is a mixed 't Hooft anomaly \cite{McGreevyGenSym}.) The effective field theory is then usually some non-linear sigma model (NLSM) for the combined Nambu-Goldstone bosons (NGBs) suppelemented by a Wess-Zumino-Witten (WZW) term, called WZW theory for short; it is generally considered unwieldy to work with, being perturbatively non-renormalizable, exacerbated by the topological term being non-local (albeit expressible as the integral of a local density in one higher dimension), its effect on the dynamics of the NGBs therefore hard to capture in a systematic calculation. Thankfully, one can often invoke IR duality to perform computations in some kind of renormalizable theory, which usually turns out to feature fermions and gauge fields. Since the resulting theory is renormalizable, one often imagines these fictitious (or ``emergent'') particles to be more elementary, or constituents of the NGBs. In fact, the canonical construction (cf., e.g., \cite{wenbook}) of non-classical ground states often proceeds in the opposite way: the fermions emerge within a so-called parton construction---essentially a particularly elaborate mean-field decoupling---of the fundamental degrees of freedom (often spins or magnetic moments living on a lattice); the gauge fields emerge, roughly speaking, from constraints needed to make the Hilbert spaces and commutation relations fit together. Such a factorization into partons is not unique; there may in fact be a whole web containing $n$ dual renormalizable theories, depending on the given setting.

Can $n$ in fact be zero? It was realized (relatively) recently \cite{StiefelPRX}, that the WZW theory governing the NGBs' low-energy dynamics might not necessarily possess a renormalizable dual description; examples were proposed with 3D spacetime as base space and the Stiefel manifold $\operatorname{St}_{N,N-4} \cong \operatorname{SO}(N)/\operatorname{SO}(4)$ as target space, for which a renormalizable dual remains unknown for $N \geqslant 7$.\footnote{The target space is in some sense a unification by generalization of the $N=5$ and $N=6$ theories. For $N=5$, the target space is the sphere $S^4$ and serves as an effective theory of SO(5) deconfined criticality \cite{senthilscience,senthilprb,Ma:2020theory,Nahum:2020note}. For $N=6$, the theory is dual to QED$_3$ and serves thus as an effective theory of the U(1) Dirac spin liquid (DSL) \cite{AffleckMarston,Hastings2000,HermeleSenthilFisher,HermeleSenthilFisher+,Song2019}.} Putative conformal phases realized by such theories thus do not admit a standard parton construction. At the same time, they may be realized in certain quasi-planar condensed matter systems, with explicit lattice constructions also proposed for $N=7$ (cf. \emph{ibid.}). Consequently, such phases, called \emph{Stiefel liquids} (SLs), are of great interest not only for intrinsic reasons, but also in the study of quantum matter; a better theoretical understanding of their underlying dynamics is hence clearly desirable. It has been argued heuristically that if $N$, which controls the intrinsic dimension of the target space, is large compared to the WZW level $k$, the theory should possess an interacting IR attractive fixed point $\operatorname{SL}^{(N,k)}$, which would represent the promised conformal phase that exists outside the realm of parton constructions. The main aim of the present work is to confront this claim with a quantitative calculation.

The absence of useful dual UV theories requires one to tackle the issue of studying the RG flow of 3D Wess-Zumino-Witten theory with $\operatorname{St}_{N,N-4}$ target space head on. Though $D=3$ is the physically relevant case, I shall largely work with the general case of $\operatorname{St}_{N,N-D-1}$ in $D$ spacetime dimensions. This has two reasons. Firstly, working in general $D$ makes the mathematical structures appearing in the theory more transparent and thereby aids intuition. Secondly, this will allow for an analytic continuation, at the level of the pertinent Feynman diagrams, of the theory to even $D$. This in turn will enable contact with the case of $D=2$, where the theory is perturbatively renormalizable by power counting (though a Lagrangian is not known). Setting $N=D+2$, such that $\operatorname{St}_{N,N-D-1} \cong S^{D+2}$ is a sphere, will recover the $\operatorname{SU}(2)_k$ Witten fixed point in $D=2$ \cite{Witten:1983ar}, which is described by level-$k$ $\operatorname{SO}(4)$ WZW theory; a (putative) generalization of this fixed point to non-spherical Stiefel manifolds is then achieved by varying $N$. In $D=2+\epsilon$ with $\epsilon$ small, there is in addition a repulsive fixed point which splits off from the Gaussian fixed point at $D=2+0^+$; if $\epsilon > \epsilon_\text{c}(k)$ is large enough, the two fixed points annihilate. The fact that $\epsilon_\text{c}(1) \approx 0.77$ is below but close to 1 (which would correspond to the physically pertinent $D=3$) explains why the putative SO(5) deconfined quantum critical point is only pseudocritical. Here, I shall derive $\epsilon_\text{c}(N,k)$ and in particular show that $\epsilon_\text{c}(N,1) > 1$ if $N \gtrsim 6$.

The fact that the $D$-dimensional theory is perturbatively non-renormalizable for $D = 3, 5, \ldots$ (i.e., where its action can be formulated) requires one to work within a ``mode decimation'' picture of RG, \`a la Wilson. Essentially, this amounts to evaluating the quantum effective action $\Gamma$ in the presence of an IR regularization $\kappa$, which acts as the renormalization scale. This way of implementing the renormalization scale has two advantages: (i) $\Gamma$ can be expanded formally in a basis of local operators (i.e., as a formal power series in fields and derivatives), the flow of couplings conversely identified by projecting $\Gamma$ onto said local operators; (ii) the change of $\Gamma$ with respect to $\kappa$ is UV-finite even if $\Gamma$ itself is not, provided the regularization is chosen judiciously. The latter is a subtle point, since the feedback of the topological WZW term (which itself cannot flow, being quantized) on the couplings of the NLSM sector of the theory become visible only at two-loop---more generally, $(D-1)$-loop---order; the way it is resolved here may be instructive in future calculations beyond the present context.

\begin{figure}
    \includegraphics[scale=1]{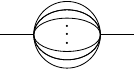}
    \caption{Feynman diagram representing the contribution of the WZW term to $\beta_g$.}
    \label{fig:Feynman}
\end{figure}

\begin{widetext}
\paragraph{Model action and setup.} The effective field theory is described by the (classical) action which reads
\begin{align}
    S &= \frac{1}{2\bar{g}} \int_{\mathbb{R}^D} \operatorname{tr}\left(\partial_\mu n^\top \partial^\mu n - \frac{2\alpha+1}{2\alpha+2} \partial_\mu n^\top n n^\top \partial^\mu n\right)d^Dx \nonumber\\
    &\hphantom{{}={}} {}+ \frac{2\pi i k}{(D+1)!(N-D-1)!\Omega_{D+1}}\int_{\mathbb{R}^D}\!\int_0^1\!\epsilon^{a_1 \ldots a_N} \epsilon^{b_1 \ldots b_{N-D-1}} \hat{n}^{a_1}_{b_1}\ldots \hat{n}^{a_{N-D-1}}_{b_{N-D-1}} \delta_{b_{N-D}b_{N-D+1}}\ldots\delta_{b_{N-1}b_{N}}\epsilon^{M_1\ldots M_{D+1}} \nonumber\\
    &\hphantom{{}={}{}+\frac{2\pi i k}{(D+1)!(N-D-1)!\Omega_D}\int_{\mathbb{R}^D}\!\int_0^1} {}\times \partial_{M_1}\hat{n}^{a_{N-D}}_{b_{N-D+1}} \ldots \partial_{M_{D+1}}\hat{n}^{a_{N}}_{b_{N}} d^Dx du.
    \label{eq:model}
\end{align}
Here, $\Omega_{D+1} = 2\pi^{(D+2)/2}/\Gamma((D+2)/2)$ is the surface area of the $(D+1)$-sphere $S^{D+1}$. The basic field $n \colon \mathbb{R}^D \to \operatorname{St}_{N,N-D-1}$ is identified as being valued in $\mathbb{R}^{N \times (N-D-1)}$ satisfying $n^\top n = I_{N-D-1}$, whilst $\hat{n} \colon \mathbb{R}^D \times [0,1] \to \operatorname{St}_{N,N-D-1}$ is an extension of $n$ to a fictitious ``bulk'' satisfying $\hat{n}(\cdot,0) = I_{N,N-D-1} \coloneqq \begin{pmatrix} I_{N-D-1} \\ O_{(D+1) \times (N-D-1)} \end{pmatrix}$ and $\hat{n}(\cdot,1) = n$; it thus (smoothly) interpolates between some reference configuration, taken to be the origin $I_{N,N-D-1} \in \operatorname{St}_{N,N-D-1}$ for concreteness, and the actual configuration $n$. The contribution of $S$ to the path integral is uniquely determined by $n = \hat{n}|_{\mathbb{R}^D \times \{ 1 \}}$ provided $k$, the so-called WZW level, is quantized to be integer.
\end{widetext}
The first line contains the non-linear sigma model with Stiefel-manifold target space.\footnote{The quartic term was neglected in \cite{StiefelPRX}, but the necessity of its existence was pointed out in previous work on the NLSM with Stiefel-manifold target space by Kunz and Zumbach \cite{KZStiefel}.} An economical way of dispensing with the constraint (which amounts to a delta function in the path integral measure) is to write the field $n$ as a Riemann exponential at the origin, thus $n = \operatorname{Exp}_{I_{N,N-D-1}}(\varphi)$, where $\varphi$ is valued in the tangent space $T_{I_{N,N-D-1}}\operatorname{St}_{N,N-D-1}$. The geometry of Stiefel manifolds has been characterized in detail in the literature. In particular, it is known, cf. e.g. \cite{zimmermann}, that $T_{I_{N,N-D-1}}\operatorname{St}_{N,N-D-1} \cong \mathfrak{so}(N-D-1) \oplus \mathbb{R}^{(D+1)\times(N-D-1)} \subset \mathbb{R}^{N\times(N-D-1)}$, such that $\varphi = \begin{pmatrix} i\mathsuet{a}_A T^A \\ \phi \end{pmatrix}$ where $T^A$ for $A = 1,\ldots,\frac12 (N-D-1)(N-D-2)$ are (Hermitean) generators of $\operatorname{SO}(N-D-1)$ (in the standard representation, unless stated otherwise). The fields $(\mathsuet{a}_A,\phi)$ are the actual dynamical degrees of freedom of the theory and can be thought of as (some generalization of) NGBs. (In particular, note that differences like $\varphi - \bar\varphi$ are well-formed objects, whilst those such as $n - \bar{n}$ are not.) Furthermore, the Riemann exponential on Stiefel manifolds can be related to matrix exponentials \cite{hmk}, which in turn can be expanded into conventional (matrix) Taylor series, leading to an equivalent Lagrangian in powers of $\mathsuet{a}$ and $\phi$. The leading contributions within the NLSM sector arise at one-loop order and requires an expansion of $S$ to quartic order in $\varphi$. (I shall, however, use the well-known relations between Ricci flow and the NLSM to derive the contributions to the beta function in that regime.) The leading vertex arising from the WZW term is a $\phi^{D+1}$-vertex containing $D$ derivatives; its contribution to the flow of $g$ is at $(D-1)$-loop order, given by the diagram Fig.~\ref{fig:Feynman}, requires a genuinely new computation.

The strategy to derive the flow equations will be to evaluate the quantum effective action $\Gamma$ (also called one-particle irreducible---1PI---effective action) in the presence of an IR cut-off $\kappa > 0$. This philosophy is hence closest to the functional renormalization group, though the details of the implementation is different.\footnote{See \cite{Gies:2006wv,kopietzIntroFRG} for textbook-level references and \cite{Dupuis:2020fhh} for a recent review.} The flow of $g$, encoded in its beta function, is fixed by the normalization of the inverse propagator of $\phi$,
\begin{align}
    -\frac{\beta_g}{g^2}\delta^D(0) &= \frac{\delta_{b_1 b_2}\delta_{c_1 c_2}}{(D+1)(N-D-1)}\frac{\delta_{\mu\nu}}{D}\nonumber\\
    & \times\left.\frac{\partial^2}{\partial p_\mu \partial p_\nu}\frac{\delta^2 \dot{\Gamma}}{\delta \phi^{c_1}_{b_1}(p)\delta \phi^{c_2}_{b_2}(-p)}\right|_{0}
\end{align}
where $(\ldots)|_{0}$ means ``vanishing fields and momenta'', the overdot denotes differentiation with respect to ``RG time'' $\ln \kappa$. The flowing action $\Gamma$ is found in perturbation theory by computing the usual 1PI Feynman diagrams, but with a modified propagator $G_0(p;\kappa)$ in place of the usual ``free'' propagator $G_0(p) = 1/p^2$. In the present case, the bottleneck is the evaluation of the WZW contribution, being of (arbitrarily) high loop order for general $D$; its sunset topology allows it still to be evaluated in closed form, but in \emph{position} space \cite{Groote:1998wy}. It hence proves more efficient to make the replacement at the level of the position-space propagator, $\hat{G}_0(x) \to \hat{G}_0(x;\kappa)$.

\paragraph{RG flow.} The beta function of $g$ works out to
\begin{align}
    \beta_g &= (D-2)g - \left[(N-2) - \frac{(N-D-2)}{2(\alpha + 1)}\right]C_1 g \nonumber\\
    &\hphantom{{}={}} {}+ \frac{C_2 (2\pi k)^2}{\Omega_{D+1}^2 D}k^2 g^{D+2},
\end{align}
where the linear term comes from making the coupling dimensionless by rescaling with suitable powers of $\kappa$, in this case $g\kappa^{D-2} \mapsto g$.
\begin{widetext}
The constants appearing above are
\begin{align}
C_1 &\coloneqq -\lim_{x\to 0} \dot{\hat{G}}_0(x;1) \\
C_2 &\coloneqq %
\det\begin{pmatrix} %
    \delta^{\mu_1}_{\mu'_1} & \ldots & \delta^{\mu_1}_{\mu'_{D-1}} \\
    \vdots & \ddots & \vdots \\
    \delta^{\mu_{D-1}}_{\mu'_1} & \ldots & \delta^{\mu_{D-1}}_{\mu'_{D-1}}
    \end{pmatrix}\left.\frac{\partial}{\partial \ln\kappa} \int_x %
    \hat{G}_{0,\mu_1}(x;\kappa)\hat{G}_{0,}^{\hphantom{0,}\mu'_1}(x;\kappa) \hat{G}_{0,\mu_{2}}^{\hphantom{0,}\mu_2'}(x;\kappa) \cdots \hat{G}_{0,\mu_{D-1}}^{\hphantom{0,}\mu'_{D-1}}(x;\kappa)\right|_{\kappa=1}
\end{align}
Both $C_{1,2} = C_{1,2r}$ are scheme-dependent for general $D>2$. At $D=2$, however, the regularization dependence drops out as expected of a dimensionless coupling at one-loop. Further setting $N=4$ reproduces the well-known beta function for 2D $\operatorname{SO}(4)_k \cong \{[\operatorname{SU}(2) \times \operatorname{SU}(2)]/\mathbb{Z}_2\}_k$ WZW theory \cite{Witten:1983ar}, as demonstrated in the Supplementary Material (SM).
\end{widetext}
The flow of the parameter $\alpha$ appearing above is fixed by the propagator of $\mathsuet{a}$. To one-loop order, it can in fact more efficiently be read off---just like the NLSM contribution to $\beta_g$---from the relation between the NLSM and Ricci flows \cite{friedan80,friedan85,CodelloPercacci}; it works out to
\begin{align}
    \beta_\alpha = \left[- \frac{N-D-3}{2} \alpha^2 + (D+1)\alpha + \frac{2D-3}{2} \right]C_1 g.
\end{align}
Further details of the derivation of the beta functions are also relegated to the SM. Fixed points of the above beta functions that are IR-stable (i.e., have no RG-relevant directions) realize scale-invariant phases.

\paragraph{Interlude: Ricci flow; geometry of $\operatorname{St}_{N,N-D-1}$.} At one-loop order, the RG flow of the NLSM sector of the theory (i.e., at vanishing WZW level $k=0$) is well-known to be given by a Ricci-type equation, $\dot{h} \propto \operatorname{Ric}_h$. The necessity of having the quartic term in Eq.~\eqref{eq:model} is seen by observing that for the induced metric, $h(\varphi,\varphi) = \frac{1}{g}\operatorname{tr} \phi^\top \phi + \frac{1}{2(1+\alpha)g}\operatorname{tr} A^\top A$ for $\varphi = \begin{pmatrix}A \\ \phi \end{pmatrix} \in T_{I_{N,N-D-1}}\operatorname{St}_{N,N-D-1}$, the point $\alpha = -\frac12$, which would make the quartic term vanish, is not a fixed point of the Ricci flow. This also makes the geometric meaning of the parameters of the theory clearer: $g$ governs the overall scaling of the metric, much like the (inverse) radius of a sphere; $\alpha$ governs the relative weight of the components of the dynamical degrees of freedom. Recall that the total number of degrees of freedom (i.e., the intrinsic dimension of $\operatorname{St}_{N,N-D-1}$) is $\frac12 (N-D-1)(N-D-2) + (D+1)(N-D-1)$. The first summand corresponds to the first summand in the identification $T_{I_{N,N-D-1}}\operatorname{St}_{N,N-D-1} \cong \mathfrak{so}(N-D-1) \oplus \mathbb{R}^{(D+1)\times(N-D-1)}$. It is only non-vanishing for $N > D+2$; for $N=D+2$, $\operatorname{St}_{D+2,1} \cong S^{D+1}$ is a sphere and automatically Einstein. Note also that the $\alpha$-contribution drops out of $\beta_g$ in this case due to the prefactor $(N-D-2)$. For $N > D+3$, there are two finite roots, given by\footnote{Note that since $\alpha$ is a dimensionless coupling, the dependence on the regularization drops out for all $D$. This will not be the case for $g$ in general.}
\begin{align}
    \alpha_*^\pm = \frac{D+1 \pm \sqrt{(2D-3)N - (D^2 + D - 10)}}{N-D-3},
\end{align}
which make $\operatorname{St}_{N,N-D-1}$ Einsteinian Riemannian manifolds. This is in accordance with the purely classical geometric considerations of \cite{Nguyen}. Among these, $\alpha_*^- \coloneqq \alpha^*_\text{SL}$ is the IR-attractive one and has the potential to describe a stable phase of matter, the putative Stiefel liquid (SL).

\paragraph{Fixed-point analysis of the WZW theory.} In what follows, I shall now focus on the IR-attractive theory subspace $\alpha = \alpha^*_\text{SL}$ and investigate whether $\beta_g \equiv \beta_g(g,\alpha^*_\text{SL})$ possesses a real IR-stable fixed point. To access the phase structure in the physical $D=3$, I shall work in $D=2+\epsilon$. Setting $D=2$ in the WZW contribution, the regularization dependence drops out, as can be anticipated from the fact that $g$ becomes dimensionless there. Thus, the beta function in $D=2+\epsilon$ is given by
\begin{align}
    \beta_g = \epsilon g - \left[(N_{D=2} - 2) - \frac{N_{D=2} - 4}{2(\alpha_{\text{s}}+1)}\right]\frac{g}{2\pi} + \frac{1}{4\pi^3}k^2 g^4.
\end{align}
Here, $g$ and $\epsilon$ are both assumed to be $O(1/k)$ $(k \to \infty)$ such that dropping all other terms $O(g^{2 \leqslant m \leqslant 4})$ $(g \to 0)$ is justified, and the analytic continuation $N \to N_D$ for $D \neq 3$ is chosen such that the second factor of the symmetry group of the theory, $\operatorname{SO}(N_D) \times \operatorname{SO}(N_D - D -1)$, is independent of $D$, mirroring standard practice in the theory of deconfined criticality \cite{Nahum:2020note,Ma:2020theory}. (In other words, $N_{D=2} = N_{D=3} - 1$; as a sanity check, the target space $\operatorname{St}_{5,1}$, which is a sphere, is analytically continued as $\operatorname{St}_{4+\epsilon,1}$, always a sphere.)

\begin{figure}
    \includegraphics[scale=.8]{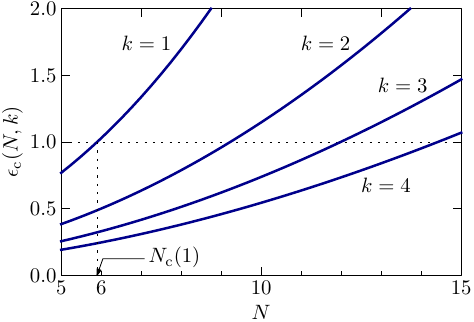}
    \caption{Critical value of $\epsilon_\text{c}(N,k)$: a real interacting fixed point in level-$k$ WZW theory with target space $\operatorname{SO}(N-1+\epsilon)/\operatorname{SO}(3+\epsilon)$ ceases to exist once the spacetime dimension exceeds $D = 2+\epsilon_\text{c}(N,k)$.}
    \label{fig:epsc}
\end{figure}

Generally, two interacting fixed points are found, located for $\alpha = \alpha^*_\text{SL}$ at
\begin{align}
    g_{*,\text{UV}} &= \frac{4 \pi  (N-2) \sqrt{N-1} \epsilon }{\sqrt{N-1} N^2-N^2-2 \sqrt{N-1} N+6 N-3 \sqrt{N-1}-5}, \\
    g_{*,\text{IR}} &= \frac{\pi}{|k|} \sqrt{\frac{N^2-\left(\sqrt{N-1}+2\right) N+5 \sqrt{N-1}-3}{N-2}} \nonumber\\
    & \hphantom{{}={}} {}+\frac{2 \pi  (N-2) \epsilon }{\left(-N+\sqrt{N-1}+2\right) N-5 \sqrt{N-1}+3} ;
\end{align}
the notation denotes whether they are attractive towards the IR or the UV. The above two formul\ae{} already illustrate the most salient feature: the separation of these two fixed points grows for increasing $N$ and decreases for increasing $\epsilon$ (and $1/k$). Once these two fixed points collide and annihilate, a stable phase of matter ceases to exist. At fixed $N$ and $k$, one can then determine the critical $\epsilon_\text{c}(N,k)$ by imposing that $\beta_g$ has a double root, i.e., $\beta_g(g_{**}) = 0$ and $\beta_g'(g_{**}) = 0$. The solution thus found for $\epsilon_\text{c}(N,k)$ is expressible analytically; the final expression, however, too cumbersome to be of practical use (apart from for $N=5$). The resulting curve for different values of $k$ is plotted in Fig.~\ref{fig:epsc} and constitutes the main result of this work.

\paragraph{Discussion.} The first interesting case is $N=5$; it has been studied extensively within the context of $\operatorname{SO}(5)$ deconfined criticality and therefore comes with a lot of prior results available to compare against. The most directly comparable computation is the Cardy-Hamber argument used by Ma and Wong \cite{Ma:2020theory}, and independetly by Nahum \cite{Nahum:2020note}. That approach yielded $\epsilon_\text{c} = \frac{4}{3\sqrt{3}}k^{-1}$; the fact that this number is nearly but just under $\epsilon =1$ means that the physics at $D=3$ is governed by a pair of complex, but nearly real, conformal field theories and famously leads to pseudocritical behavior. The present computation reproduces this $\epsilon_\text{c}$ exactly and provides a non-trivial check of the prescription proposed herein for analytic continuation of the WZW contribution with Stiefel-manifolds target space to general $D$. (Recall that the theory cannot be defined, at least in the conventional sense by a Lagrangian, if $D$ is not odd.) Given that the $k=1$ theory for $N=5$ appears to only feature complex interacting fixed points, the natural question to ask then is how large $N$ has to be to support real interacting fixed points $\operatorname{SL}^{(N,k)}$ in $D=3$. This turns out to already be the case for $N=6$, for good reason. It is known that 3D level-$1$ WZW theory with target space $\operatorname{St}_{6,2}$ is dual to QED$_3$ with $N_\text{f}=4$ fermion flavors \cite{StiefelPRX}, which simulations have suggested to be conformal \cite{karthiknarayanan1,karthiknarayanan2}.\footnote{The critical flavor number is a matter of debate, but there is consensus on it being safely below $N_\text{f}=4$.} Within the present formulation, one can also readily compute the scaling dimension of the lowest scalar; it is given approximately by $\Delta_\text{S} = \beta_g'(g_{*,\text{IR}}) \approx 3.3$. One can also define a critical $N_\text{c}(k)$ implicitly by $\epsilon_\text{c}(N_\text{c}(k),k) = 1$. It turns out that $N_\text{c}(1) \approx 5.9$; the fact that this number is quite close to $6$ is not unexpected: were it much below $6$, $\Delta_\text{S}$ would exceed the bound $\Delta_\text{S} < 4$, which is required of any QED$_3$ theory. As to the precise value of $\Delta_\text{S}$, to next-to-leading order (NLO) in large $N_\text{f}$, the same dimension was calculated to be $\Delta_\text{S} = 4 - (\sqrt{7}-2)/(3\pi^2 N_\text{f}) + O(1/N_\text{f}^2)$ by Xu \cite{XuQED3} using RG techniques and by Chester and Pufu \cite{ChesterPufuQED3} using (analytical) conformal bootstrap, whence $\Delta_\text{S} \approx 3.65$. Meanwhile, exploiting the fact that QED$_D$ becomes perturbatively renormalizable in $D=4$, di Pietro \emph{et al.} \cite{diPietroQEDeps} derived in $4-2\varepsilon$ expansion $\Delta_\text{S} = 6-4\varepsilon + 3 \varepsilon\left(1+2N_\text{f} - 2\sqrt{N_\text{f}^2 + N_\text{f} + 25}\right)/(4N_\text{f}) + O(\varepsilon^2)$, whence $\Delta_\text{S} \approx 3.58$. Though the present estimate from WZW theory in $D=2+\epsilon$ is systematically lower than the complementary estimates from the dual QED$_3$ theory, it is still remarkably close (within 5-10\%) given the very different nature of the operators involved.

For $N=7$, a stable real fixed point exists \emph{a fortiori}, and the scaling dimension of the lowest singlet works out to $\Delta_\text{S} \approx 5.2$. To the best of my knowledge, this is the first concrete theoretical prediction of a scaling dimension for a WZW theory with no known gauge theory dual. The mechanism for $\Delta_\text{S}$ being significantly greater than four may be understood as follows: the WZW contribution to $\beta_g$ does not grow with $N$. Consequently, for fixed $k$, $\epsilon_\text{c}(N,k)$ increases with $N$. Since $\Delta_\text{S} - d \propto \sqrt{(d-2) - \epsilon_\text{c}(N,k)}$,\footnote{In other words, the slope of the beta function at the stable fixed point increases with $N$. This weaker version of the statement is actually sufficient to establish the veracity of the assertion regarding $\Delta_\text{S}$.} ultimately, $\Delta_\text{S}$ will exceed 4 for large enough $N$. Meanwhile, for QCD$_3$-like theories, one typically expects $\Delta_\text{S} < 4$. Should such Stiefel-WZW theories possess a (super-)renormalizable gauge dual, it would have to be exotic. It is worth contrasting the situation with the case of level-$k$ WZW theory with Grassmannian target manifold $\operatorname{SU}(2N)/\operatorname{SU}(N)^2$, indeed dual to $(2N)$-flavor QCD$_3$ with gauge group $\operatorname{SU}(k)$ \cite{KomargodskiSeibergQCD3}. There, the contribution of the WZW term scales with $N$ \cite{BietalGrassmann}, which would allow $\Delta_\text{S}$ to remain within the bound also for large $N$.

\paragraph{Outlook.}

There are some subtleties regarding this computation which warrant closer look and present scope for future investigation. First, there is the question of higher-order corrections. It is important, because like in any $\epsilon$ expansion, ultimately an extrapolation to $\epsilon = 1$ is required. This will presumably be a variant of the loop expansion, though the power counting needs to be modified to account for the fact that it is a dual expansion in $\epsilon$ and $1/k$ (with $k=1$ again being the physically most important case). An alternative may be to work in fixed dimension but at large $N$ and $k$. Note that the present one-loop calculation appears to suggest that the NLSM fixed point to leading order becomes weakly interacting at fixed $D$, if $N$ is large. However, power counting in the NLSM is non-trivial in the large $N$ limit, amounting essentially to a genus expansion, with planar diagrams being the lowest-order expansion. For unitary matrix models, this has been done using Dyson-Schwinger techniques \cite{CampostriniRossiVicari}, which may be adaptible to the present context. Alternatively, a complementary approach using holographic techniques may be worth attempting. A completely orthogonal approach to setting up an $\epsilon$ expansion may be to consider long-ranged deformations of the WZW theory. This may be deemed promising in view of recent progress towards renormalizing the long-ranged NLSM model \cite{Gubser:2019uyf}, provided the techniques can be adapted to the presence of the WZW term accordingly. Yet another approach would be to eschew control parameters altogether, and work non-perturbatively using (some variant of) the functional renormalization group. This was done recently for $N=5$ \cite{RayNatCommun}, and there appear to be no obvious obstructions to adapting those techniques to other values of $N$; the diagram topology though may be a bit more involved, at least compared to those treated in this paper, due to the higher-order regulator insertions. An intermediate approach may comprise evaluating the expressions directly in $D=3$, though it would be uncontrolled. A theoretical challenge would then lie in tackling the regulator dependence---my initial attempts in this direction show that the regulator dependence is rather severe, and exacerbated by the fact that a priori means of finding the optimal regulator, e.g., the gap criterion, do not apply when the regularization is implemented as a deformation of the real-space propagator. These considerations may hence also be of independent intrinsic interest to practitioners of non-perturbative approaches to the renormalization group. %That must, hence, remain a different story for another time.

\paragraph{Acknowledgments.} I thank A. Eichhorn, B. Hawashin, L. Janssen and M.~M. Scherer for enlightening discussions and collaboration on related work. Further useful discussions with S. Bhattacharya, A.~A. Christensen, R. Gurau, Y.-C. He, P. Marquard, J.~M. Pawlowski, M.~J. Thomsen and C. Wang are also gratefully acknowledged. %
%Support from are gratefully acknowledged.

%%%%%%%%%%%%%%%%%%%%%%%%%%%%%%%%%%%%%%%%%%%%%%%%%%%%%%%%%%%%%%%%%%%%%%%
% BIBLIOGRAPHY: FOR USE WITH BIBTEX
%%%%%%%%%%%%%%%%%%%%%%%%%%%%%%%%%%%%%%%%%%%%%%%%%%%%%%%%%%%%%%%%%%%%%%%
% \bibliographystyle{apsrev4-2}
% \bibliography{Stiefel}

\begin{thebibliography}{36}%
\makeatletter
\providecommand \@ifxundefined [1]{%
 \@ifx{#1\undefined}
}%
\providecommand \@ifnum [1]{%
 \ifnum #1\expandafter \@firstoftwo
 \else \expandafter \@secondoftwo
 \fi
}%
\providecommand \@ifx [1]{%
 \ifx #1\expandafter \@firstoftwo
 \else \expandafter \@secondoftwo
 \fi
}%
\providecommand \natexlab [1]{#1}%
\providecommand \enquote  [1]{``#1''}%
\providecommand \bibnamefont  [1]{#1}%
\providecommand \bibfnamefont [1]{#1}%
\providecommand \citenamefont [1]{#1}%
\providecommand \href@noop [0]{\@secondoftwo}%
\providecommand \href [0]{\begingroup \@sanitize@url \@href}%
\providecommand \@href[1]{\@@startlink{#1}\@@href}%
\providecommand \@@href[1]{\endgroup#1\@@endlink}%
\providecommand \@sanitize@url [0]{\catcode `\\12\catcode `\$12\catcode
  `\&12\catcode `\#12\catcode `\^12\catcode `\_12\catcode `\%12\relax}%
\providecommand \@@startlink[1]{}%
\providecommand \@@endlink[0]{}%
\providecommand \url  [0]{\begingroup\@sanitize@url \@url }%
\providecommand \@url [1]{\endgroup\@href {#1}{\urlprefix }}%
\providecommand \urlprefix  [0]{URL }%
\providecommand \Eprint [0]{\href }%
\providecommand \doibase [0]{https://doi.org/}%
\providecommand \selectlanguage [0]{\@gobble}%
\providecommand \bibinfo  [0]{\@secondoftwo}%
\providecommand \bibfield  [0]{\@secondoftwo}%
\providecommand \translation [1]{[#1]}%
\providecommand \BibitemOpen [0]{}%
\providecommand \bibitemStop [0]{}%
\providecommand \bibitemNoStop [0]{.\EOS\space}%
\providecommand \EOS [0]{\spacefactor3000\relax}%
\providecommand \BibitemShut  [1]{\csname bibitem#1\endcsname}%
\let\auto@bib@innerbib\@empty
%</preamble>
\bibitem [{\citenamefont {Senthil}\ and\ \citenamefont
  {Fisher}(2006)}]{Senthil:2005jk}%
  \BibitemOpen
  \bibfield  {author} {\bibinfo {author} {\bibfnamefont {T.}~\bibnamefont
  {Senthil}}\ and\ \bibinfo {author} {\bibfnamefont {M.~P.~A.}\ \bibnamefont
  {Fisher}},\ }\href {https://doi.org/10.1103/PhysRevB.74.064405} {\bibfield
  {journal} {\bibinfo  {journal} {Phys. Rev. B}\ }\textbf {\bibinfo {volume}
  {74}},\ \bibinfo {pages} {064405} (\bibinfo {year} {2006})},\ \Eprint
  {https://arxiv.org/abs/cond-mat/0510459} {arXiv:cond-mat/0510459}
  \BibitemShut {NoStop}%
\bibitem [{\citenamefont {Fradkin}\ \emph {et~al.}(2015)\citenamefont
  {Fradkin}, \citenamefont {Kivelson},\ and\ \citenamefont
  {Tranquada}}]{FradkinColloquium}%
  \BibitemOpen
  \bibfield  {author} {\bibinfo {author} {\bibfnamefont {E.}~\bibnamefont
  {Fradkin}}, \bibinfo {author} {\bibfnamefont {S.~A.}\ \bibnamefont
  {Kivelson}},\ and\ \bibinfo {author} {\bibfnamefont {J.~M.}\ \bibnamefont
  {Tranquada}},\ }\href {https://doi.org/10.1103/RevModPhys.87.457} {\bibfield
  {journal} {\bibinfo  {journal} {Rev. Mod. Phys.}\ }\textbf {\bibinfo {volume}
  {87}},\ \bibinfo {pages} {457} (\bibinfo {year} {2015})}\BibitemShut
  {NoStop}%
\bibitem [{\citenamefont {McGreevy}(2023)}]{McGreevyGenSym}%
  \BibitemOpen
  \bibfield  {author} {\bibinfo {author} {\bibfnamefont {J.}~\bibnamefont
  {McGreevy}},\ }\href
  {https://doi.org/10.1146/annurev-conmatphys-040721-021029} {\bibfield
  {journal} {\bibinfo  {journal} {Ann. Rev. Condensed Matter Phys.}\ }\textbf
  {\bibinfo {volume} {14}},\ \bibinfo {pages} {57} (\bibinfo {year}
  {2023})}\BibitemShut {NoStop}%
\bibitem [{\citenamefont {Wen}(2004)}]{wenbook}%
  \BibitemOpen
  \bibfield  {author} {\bibinfo {author} {\bibfnamefont {X.-G.}\ \bibnamefont
  {Wen}},\ }\href@noop {} {\emph {\bibinfo {title} {Quantum field theory of
  many-body systems: From the origin of sound to an origin of light and
  electrons}}}\ (\bibinfo  {publisher} {Oxford university press},\ \bibinfo
  {year} {2004})\BibitemShut {NoStop}%
\bibitem [{\citenamefont {Zou}\ \emph {et~al.}(2021)\citenamefont {Zou},
  \citenamefont {He},\ and\ \citenamefont {Wang}}]{StiefelPRX}%
  \BibitemOpen
  \bibfield  {author} {\bibinfo {author} {\bibfnamefont {L.}~\bibnamefont
  {Zou}}, \bibinfo {author} {\bibfnamefont {Y.-C.}\ \bibnamefont {He}},\ and\
  \bibinfo {author} {\bibfnamefont {C.}~\bibnamefont {Wang}},\ }\href@noop {}
  {\bibfield  {journal} {\bibinfo  {journal} {Phys. Rev. X}\ }\textbf {\bibinfo
  {volume} {11}},\ \bibinfo {pages} {031043} (\bibinfo {year}
  {2021})}\BibitemShut {NoStop}%
\bibitem [{\citenamefont {Senthil}\ \emph
  {et~al.}(2004{\natexlab{a}})\citenamefont {Senthil}, \citenamefont
  {Vishwanath}, \citenamefont {Balents}, \citenamefont {Sachdev},\ and\
  \citenamefont {Fisher}}]{senthilscience}%
  \BibitemOpen
  \bibfield  {author} {\bibinfo {author} {\bibfnamefont {T.}~\bibnamefont
  {Senthil}}, \bibinfo {author} {\bibfnamefont {A.}~\bibnamefont {Vishwanath}},
  \bibinfo {author} {\bibfnamefont {L.}~\bibnamefont {Balents}}, \bibinfo
  {author} {\bibfnamefont {S.}~\bibnamefont {Sachdev}},\ and\ \bibinfo {author}
  {\bibfnamefont {M.~P.~A.}\ \bibnamefont {Fisher}},\ }\href
  {https://doi.org/10.1126/science.1091806} {\bibfield  {journal} {\bibinfo
  {journal} {Science}\ }\textbf {\bibinfo {volume} {303}},\ \bibinfo {pages}
  {1490} (\bibinfo {year} {2004}{\natexlab{a}})}\BibitemShut {NoStop}%
\bibitem [{\citenamefont {Senthil}\ \emph
  {et~al.}(2004{\natexlab{b}})\citenamefont {Senthil}, \citenamefont {Balents},
  \citenamefont {Sachdev}, \citenamefont {Vishwanath},\ and\ \citenamefont
  {Fisher}}]{senthilprb}%
  \BibitemOpen
  \bibfield  {author} {\bibinfo {author} {\bibfnamefont {T.}~\bibnamefont
  {Senthil}}, \bibinfo {author} {\bibfnamefont {L.}~\bibnamefont {Balents}},
  \bibinfo {author} {\bibfnamefont {S.}~\bibnamefont {Sachdev}}, \bibinfo
  {author} {\bibfnamefont {A.}~\bibnamefont {Vishwanath}},\ and\ \bibinfo
  {author} {\bibfnamefont {M.~P.~A.}\ \bibnamefont {Fisher}},\ }\href
  {https://doi.org/10.1103/PhysRevB.70.144407} {\bibfield  {journal} {\bibinfo
  {journal} {Phys. Rev. B}\ }\textbf {\bibinfo {volume} {70}},\ \bibinfo
  {pages} {144407} (\bibinfo {year} {2004}{\natexlab{b}})}\BibitemShut
  {NoStop}%
\bibitem [{\citenamefont {Ma}\ and\ \citenamefont
  {Wang}(2020)}]{Ma:2020theory}%
  \BibitemOpen
  \bibfield  {author} {\bibinfo {author} {\bibfnamefont {R.}~\bibnamefont
  {Ma}}\ and\ \bibinfo {author} {\bibfnamefont {C.}~\bibnamefont {Wang}},\
  }\href {https://doi.org/10.1103/PhysRevB.102.020407} {\bibfield  {journal}
  {\bibinfo  {journal} {Phys. Rev. B}\ }\textbf {\bibinfo {volume} {102}},\
  \bibinfo {pages} {020407} (\bibinfo {year} {2020})}\BibitemShut {NoStop}%
\bibitem [{\citenamefont {Nahum}(2020)}]{Nahum:2020note}%
  \BibitemOpen
  \bibfield  {author} {\bibinfo {author} {\bibfnamefont {A.}~\bibnamefont
  {Nahum}},\ }\href {https://doi.org/10.1103/PhysRevB.102.201116} {\bibfield
  {journal} {\bibinfo  {journal} {Phys. Rev. B}\ }\textbf {\bibinfo {volume}
  {102}},\ \bibinfo {pages} {201116} (\bibinfo {year} {2020})}\BibitemShut
  {NoStop}%
\bibitem [{\citenamefont {Affleck}\ and\ \citenamefont
  {Marston}(1988)}]{AffleckMarston}%
  \BibitemOpen
  \bibfield  {author} {\bibinfo {author} {\bibfnamefont {I.}~\bibnamefont
  {Affleck}}\ and\ \bibinfo {author} {\bibfnamefont {J.~B.}\ \bibnamefont
  {Marston}},\ }\href {https://doi.org/10.1103/PhysRevB.37.3774} {\bibfield
  {journal} {\bibinfo  {journal} {Phys. Rev. B}\ }\textbf {\bibinfo {volume}
  {37}},\ \bibinfo {pages} {3774} (\bibinfo {year} {1988})}\BibitemShut
  {NoStop}%
\bibitem [{\citenamefont {Hastings}(2000)}]{Hastings2000}%
  \BibitemOpen
  \bibfield  {author} {\bibinfo {author} {\bibfnamefont {M.~B.}\ \bibnamefont
  {Hastings}},\ }\href {https://doi.org/10.1103/PhysRevB.63.014413} {\bibfield
  {journal} {\bibinfo  {journal} {Phys. Rev. B}\ }\textbf {\bibinfo {volume}
  {63}},\ \bibinfo {pages} {014413} (\bibinfo {year} {2000})}\BibitemShut
  {NoStop}%
\bibitem [{\citenamefont {Hermele}\ \emph {et~al.}(2005)\citenamefont
  {Hermele}, \citenamefont {Senthil},\ and\ \citenamefont
  {Fisher}}]{HermeleSenthilFisher}%
  \BibitemOpen
  \bibfield  {author} {\bibinfo {author} {\bibfnamefont {M.}~\bibnamefont
  {Hermele}}, \bibinfo {author} {\bibfnamefont {T.}~\bibnamefont {Senthil}},\
  and\ \bibinfo {author} {\bibfnamefont {M.~P.~A.}\ \bibnamefont {Fisher}},\
  }\href {https://doi.org/10.1103/PhysRevB.72.104404} {\bibfield  {journal}
  {\bibinfo  {journal} {Phys. Rev. B}\ }\textbf {\bibinfo {volume} {72}},\
  \bibinfo {pages} {104404} (\bibinfo {year} {2005})}\BibitemShut {NoStop}%
\bibitem [{\citenamefont {Hermele}\ \emph {et~al.}(2004)\citenamefont
  {Hermele}, \citenamefont {Senthil}, \citenamefont {Fisher}, \citenamefont
  {Lee}, \citenamefont {Nagaosa},\ and\ \citenamefont
  {Wen}}]{HermeleSenthilFisher+}%
  \BibitemOpen
  \bibfield  {author} {\bibinfo {author} {\bibfnamefont {M.}~\bibnamefont
  {Hermele}}, \bibinfo {author} {\bibfnamefont {T.}~\bibnamefont {Senthil}},
  \bibinfo {author} {\bibfnamefont {M.~P.~A.}\ \bibnamefont {Fisher}}, \bibinfo
  {author} {\bibfnamefont {P.~A.}\ \bibnamefont {Lee}}, \bibinfo {author}
  {\bibfnamefont {N.}~\bibnamefont {Nagaosa}},\ and\ \bibinfo {author}
  {\bibfnamefont {X.-G.}\ \bibnamefont {Wen}},\ }\href
  {https://doi.org/10.1103/PhysRevB.70.214437} {\bibfield  {journal} {\bibinfo
  {journal} {Phys. Rev. B}\ }\textbf {\bibinfo {volume} {70}},\ \bibinfo
  {pages} {214437} (\bibinfo {year} {2004})}\BibitemShut {NoStop}%
\bibitem [{\citenamefont {Song}\ \emph {et~al.}(2019)\citenamefont {Song},
  \citenamefont {Wang}, \citenamefont {Vishwanath},\ and\ \citenamefont
  {He}}]{Song2019}%
  \BibitemOpen
  \bibfield  {author} {\bibinfo {author} {\bibfnamefont {X.-Y.}\ \bibnamefont
  {Song}}, \bibinfo {author} {\bibfnamefont {C.}~\bibnamefont {Wang}}, \bibinfo
  {author} {\bibfnamefont {A.}~\bibnamefont {Vishwanath}},\ and\ \bibinfo
  {author} {\bibfnamefont {Y.-C.}\ \bibnamefont {He}},\ }\href
  {https://doi.org/10.1038/s41467-019-11727-3} {\bibfield  {journal} {\bibinfo
  {journal} {Nat. Commun.}\ }\textbf {\bibinfo {volume} {10}},\ \bibinfo
  {pages} {4254} (\bibinfo {year} {2019})}\BibitemShut {NoStop}%
\bibitem [{\citenamefont {Witten}(1984)}]{Witten:1983ar}%
  \BibitemOpen
  \bibfield  {author} {\bibinfo {author} {\bibfnamefont {E.}~\bibnamefont
  {Witten}},\ }\href {https://doi.org/10.1007/BF01215276} {\bibfield  {journal}
  {\bibinfo  {journal} {Commun. Math. Phys.}\ }\textbf {\bibinfo {volume}
  {92}},\ \bibinfo {pages} {455} (\bibinfo {year} {1984})}\BibitemShut
  {NoStop}%
\bibitem [{\citenamefont {Kunz}\ and\ \citenamefont
  {Zumbach}(1993)}]{KZStiefel}%
  \BibitemOpen
  \bibfield  {author} {\bibinfo {author} {\bibfnamefont {H.}~\bibnamefont
  {Kunz}}\ and\ \bibinfo {author} {\bibfnamefont {G.}~\bibnamefont {Zumbach}},\
  }\href@noop {} {\bibfield  {journal} {\bibinfo  {journal} {J. Phys. A: Math.
  Gen.}\ }\textbf {\bibinfo {volume} {26}},\ \bibinfo {pages} {3121} (\bibinfo
  {year} {1993})}\BibitemShut {NoStop}%
\bibitem [{\citenamefont {Zimmermann}\ and\ \citenamefont
  {H\"{u}per}(2022)}]{zimmermann}%
  \BibitemOpen
  \bibfield  {author} {\bibinfo {author} {\bibfnamefont {R.}~\bibnamefont
  {Zimmermann}}\ and\ \bibinfo {author} {\bibfnamefont {K.}~\bibnamefont
  {H\"{u}per}},\ }\href {https://doi.org/10.1137/21M1425426} {\bibfield
  {journal} {\bibinfo  {journal} {SIAM J. Matrix Anal. Appl.}\ }\textbf
  {\bibinfo {volume} {43}},\ \bibinfo {pages} {953} (\bibinfo {year}
  {2022})}\BibitemShut {NoStop}%
\bibitem [{\citenamefont {Knut}\ \emph {et~al.}(2021)\citenamefont {Knut},
  \citenamefont {Irina},\ and\ \citenamefont {Silva}}]{hmk}%
  \BibitemOpen
  \bibfield  {author} {\bibinfo {author} {\bibfnamefont {H.}~\bibnamefont
  {Knut}}, \bibinfo {author} {\bibfnamefont {M.}~\bibnamefont {Irina}},\ and\
  \bibinfo {author} {\bibfnamefont {L.~F.}\ \bibnamefont {Silva}},\ }\href
  {https://doi.org/10.3934/jgm.2020031} {\bibfield  {journal} {\bibinfo
  {journal} {J. Geom. Mech.}\ }\textbf {\bibinfo {volume} {13}},\ \bibinfo
  {pages} {55} (\bibinfo {year} {2021})}\BibitemShut {NoStop}%
\bibitem [{\citenamefont {Gies}(2012)}]{Gies:2006wv}%
  \BibitemOpen
  \bibfield  {author} {\bibinfo {author} {\bibfnamefont {H.}~\bibnamefont
  {Gies}},\ }\href {https://doi.org/10.1007/978-3-642-27320-9_6} {\bibfield
  {journal} {\bibinfo  {journal} {Lect. Notes Phys.}\ }\textbf {\bibinfo
  {volume} {852}},\ \bibinfo {pages} {287} (\bibinfo {year}
  {2012})}\BibitemShut {NoStop}%
\bibitem [{\citenamefont {Kopietz}\ \emph {et~al.}(2010)\citenamefont
  {Kopietz}, \citenamefont {Bartosch},\ and\ \citenamefont
  {Sch{\"u}tz}}]{kopietzIntroFRG}%
  \BibitemOpen
  \bibfield  {author} {\bibinfo {author} {\bibfnamefont {P.}~\bibnamefont
  {Kopietz}}, \bibinfo {author} {\bibfnamefont {L.}~\bibnamefont {Bartosch}},\
  and\ \bibinfo {author} {\bibfnamefont {F.}~\bibnamefont {Sch{\"u}tz}},\
  }\href@noop {} {\emph {\bibinfo {title} {Introduction to the functional
  renormalization group}}}\ (\bibinfo  {publisher} {Springer Science \&
  Business Media},\ \bibinfo {year} {2010})\BibitemShut {NoStop}%
\bibitem [{\citenamefont {Dupuis}\ \emph {et~al.}(2021)\citenamefont {Dupuis},
  \citenamefont {Canet}, \citenamefont {Eichhorn}, \citenamefont {Metzner},
  \citenamefont {Pawlowski}, \citenamefont {Tissier},\ and\ \citenamefont
  {Wschebor}}]{Dupuis:2020fhh}%
  \BibitemOpen
  \bibfield  {author} {\bibinfo {author} {\bibfnamefont {N.}~\bibnamefont
  {Dupuis}}, \bibinfo {author} {\bibfnamefont {L.}~\bibnamefont {Canet}},
  \bibinfo {author} {\bibfnamefont {A.}~\bibnamefont {Eichhorn}}, \bibinfo
  {author} {\bibfnamefont {W.}~\bibnamefont {Metzner}}, \bibinfo {author}
  {\bibfnamefont {J.~M.}\ \bibnamefont {Pawlowski}}, \bibinfo {author}
  {\bibfnamefont {M.}~\bibnamefont {Tissier}},\ and\ \bibinfo {author}
  {\bibfnamefont {N.}~\bibnamefont {Wschebor}},\ }\href
  {https://doi.org/10.1016/j.physrep.2021.01.001} {\bibfield  {journal}
  {\bibinfo  {journal} {Phys. Rept.}\ }\textbf {\bibinfo {volume} {910}},\
  \bibinfo {pages} {1} (\bibinfo {year} {2021})}\BibitemShut {NoStop}%
\bibitem [{\citenamefont {Groote}\ \emph {et~al.}(1999)\citenamefont {Groote},
  \citenamefont {Korner},\ and\ \citenamefont {Pivovarov}}]{Groote:1998wy}%
  \BibitemOpen
  \bibfield  {author} {\bibinfo {author} {\bibfnamefont {S.}~\bibnamefont
  {Groote}}, \bibinfo {author} {\bibfnamefont {J.~G.}\ \bibnamefont {Korner}},\
  and\ \bibinfo {author} {\bibfnamefont {A.~A.}\ \bibnamefont {Pivovarov}},\
  }\href {https://doi.org/10.1016/S0550-3213(98)00812-8} {\bibfield  {journal}
  {\bibinfo  {journal} {Nucl. Phys. B}\ }\textbf {\bibinfo {volume} {542}},\
  \bibinfo {pages} {515} (\bibinfo {year} {1999})},\ \Eprint
  {https://arxiv.org/abs/hep-ph/9806402} {arXiv:hep-ph/9806402} \BibitemShut
  {NoStop}%
\bibitem [{\citenamefont {Friedan}(1980)}]{friedan80}%
  \BibitemOpen
  \bibfield  {author} {\bibinfo {author} {\bibfnamefont {D.}~\bibnamefont
  {Friedan}},\ }\href {https://doi.org/10.1103/PhysRevLett.45.1057} {\bibfield
  {journal} {\bibinfo  {journal} {Phys. Rev. Lett.}\ }\textbf {\bibinfo
  {volume} {45}},\ \bibinfo {pages} {1057} (\bibinfo {year}
  {1980})}\BibitemShut {NoStop}%
\bibitem [{\citenamefont {Friedan}(1985)}]{friedan85}%
  \BibitemOpen
  \bibfield  {author} {\bibinfo {author} {\bibfnamefont {D.~H.}\ \bibnamefont
  {Friedan}},\ }\href
  {https://doi.org/https://doi.org/10.1016/0003-4916(85)90384-7} {\bibfield
  {journal} {\bibinfo  {journal} {Ann. Phys. (N. Y.)}\ }\textbf {\bibinfo
  {volume} {163}},\ \bibinfo {pages} {318} (\bibinfo {year}
  {1985})}\BibitemShut {NoStop}%
\bibitem [{\citenamefont {Codello}\ and\ \citenamefont
  {Percacci}(2009)}]{CodelloPercacci}%
  \BibitemOpen
  \bibfield  {author} {\bibinfo {author} {\bibfnamefont {A.}~\bibnamefont
  {Codello}}\ and\ \bibinfo {author} {\bibfnamefont {R.}~\bibnamefont
  {Percacci}},\ }\href
  {https://doi.org/https://doi.org/10.1016/j.physletb.2009.01.032} {\bibfield
  {journal} {\bibinfo  {journal} {Phys. Lett. B}\ }\textbf {\bibinfo {volume}
  {672}},\ \bibinfo {pages} {280} (\bibinfo {year} {2009})}\BibitemShut
  {NoStop}%
\bibitem [{\citenamefont {Nguyen}(2021)}]{Nguyen}%
  \BibitemOpen
  \bibfield  {author} {\bibinfo {author} {\bibfnamefont {D.}~\bibnamefont
  {Nguyen}},\ }\href {https://doi.org/10.48550/arXiv.2105.01834} {\bibfield
  {journal} {\bibinfo  {journal} {J. Lie Theory}\ }\textbf {\bibinfo {volume}
  {32}},\ \bibinfo {pages} {563} (\bibinfo {year} {2021})}\BibitemShut
  {NoStop}%
\bibitem [{\citenamefont {Karthik}\ and\ \citenamefont
  {Narayanan}(2016{\natexlab{a}})}]{karthiknarayanan1}%
  \BibitemOpen
  \bibfield  {author} {\bibinfo {author} {\bibfnamefont {N.}~\bibnamefont
  {Karthik}}\ and\ \bibinfo {author} {\bibfnamefont {R.}~\bibnamefont
  {Narayanan}},\ }\href {https://doi.org/10.1103/PhysRevD.94.065026} {\bibfield
   {journal} {\bibinfo  {journal} {Phys. Rev. D}\ }\textbf {\bibinfo {volume}
  {94}},\ \bibinfo {pages} {065026} (\bibinfo {year}
  {2016}{\natexlab{a}})}\BibitemShut {NoStop}%
\bibitem [{\citenamefont {Karthik}\ and\ \citenamefont
  {Narayanan}(2016{\natexlab{b}})}]{karthiknarayanan2}%
  \BibitemOpen
  \bibfield  {author} {\bibinfo {author} {\bibfnamefont {N.}~\bibnamefont
  {Karthik}}\ and\ \bibinfo {author} {\bibfnamefont {R.}~\bibnamefont
  {Narayanan}},\ }\href {https://doi.org/10.1103/PhysRevD.93.045020} {\bibfield
   {journal} {\bibinfo  {journal} {Phys. Rev. D}\ }\textbf {\bibinfo {volume}
  {93}},\ \bibinfo {pages} {045020} (\bibinfo {year}
  {2016}{\natexlab{b}})}\BibitemShut {NoStop}%
\bibitem [{\citenamefont {Xu}(2008)}]{XuQED3}%
  \BibitemOpen
  \bibfield  {author} {\bibinfo {author} {\bibfnamefont {C.}~\bibnamefont
  {Xu}},\ }\href {https://doi.org/10.1103/PhysRevB.78.054432} {\bibfield
  {journal} {\bibinfo  {journal} {Phys. Rev. B}\ }\textbf {\bibinfo {volume}
  {78}},\ \bibinfo {pages} {054432} (\bibinfo {year} {2008})}\BibitemShut
  {NoStop}%
\bibitem [{\citenamefont {Chester}\ and\ \citenamefont
  {Pufu}(2016)}]{ChesterPufuQED3}%
  \BibitemOpen
  \bibfield  {author} {\bibinfo {author} {\bibfnamefont {S.~M.}\ \bibnamefont
  {Chester}}\ and\ \bibinfo {author} {\bibfnamefont {S.~S.}\ \bibnamefont
  {Pufu}},\ }\href {https://doi.org/10.1007/JHEP08(2016)069} {\bibfield
  {journal} {\bibinfo  {journal} {J. High Energy Phys.}\ }\textbf {\bibinfo
  {volume} {2016}}\bibinfo  {number} { (8)},\ \bibinfo {pages}
  {69}}\BibitemShut {NoStop}%
\bibitem [{\citenamefont {Di~Pietro}\ \emph {et~al.}(2016)\citenamefont
  {Di~Pietro}, \citenamefont {Komargodski}, \citenamefont {Shamir},\ and\
  \citenamefont {Stamou}}]{diPietroQEDeps}%
  \BibitemOpen
\bibfield  {number} {  }\bibfield  {author} {\bibinfo {author} {\bibfnamefont
  {L.}~\bibnamefont {Di~Pietro}}, \bibinfo {author} {\bibfnamefont
  {Z.}~\bibnamefont {Komargodski}}, \bibinfo {author} {\bibfnamefont
  {I.}~\bibnamefont {Shamir}},\ and\ \bibinfo {author} {\bibfnamefont
  {E.}~\bibnamefont {Stamou}},\ }\href
  {https://doi.org/10.1103/PhysRevLett.116.131601} {\bibfield  {journal}
  {\bibinfo  {journal} {Phys. Rev. Lett.}\ }\textbf {\bibinfo {volume} {116}},\
  \bibinfo {pages} {131601} (\bibinfo {year} {2016})}\BibitemShut {NoStop}%
\bibitem [{\citenamefont {Komargodski}\ and\ \citenamefont
  {Seiberg}(2018)}]{KomargodskiSeibergQCD3}%
  \BibitemOpen
  \bibfield  {author} {\bibinfo {author} {\bibfnamefont {Z.}~\bibnamefont
  {Komargodski}}\ and\ \bibinfo {author} {\bibfnamefont {N.}~\bibnamefont
  {Seiberg}},\ }\href {https://doi.org/10.1007/JHEP01(2018)109} {\bibfield
  {journal} {\bibinfo  {journal} {J. High Energy Phys.}\ }\textbf {\bibinfo
  {volume} {2018}}\bibinfo  {number} { (1)},\ \bibinfo {pages}
  {109}}\BibitemShut {NoStop}%
\bibitem [{\citenamefont {Bi}\ \emph {et~al.}(2016)\citenamefont {Bi},
  \citenamefont {Rasmussen}, \citenamefont {BenTov},\ and\ \citenamefont
  {Xu}}]{BietalGrassmann}%
  \BibitemOpen
\bibfield  {number} {  }\bibfield  {author} {\bibinfo {author} {\bibfnamefont
  {Z.}~\bibnamefont {Bi}}, \bibinfo {author} {\bibfnamefont {A.}~\bibnamefont
  {Rasmussen}}, \bibinfo {author} {\bibfnamefont {Y.}~\bibnamefont {BenTov}},\
  and\ \bibinfo {author} {\bibfnamefont {C.}~\bibnamefont {Xu}},\ }\href@noop
  {} {\bibfield  {journal} {\bibinfo  {journal} {arXiv:1605.05336}\ } (\bibinfo
  {year} {2016})}\BibitemShut {NoStop}%
\bibitem [{\citenamefont {Campostrini}\ \emph {et~al.}(1995)\citenamefont
  {Campostrini}, \citenamefont {Rossi},\ and\ \citenamefont
  {Vicari}}]{CampostriniRossiVicari}%
  \BibitemOpen
  \bibfield  {author} {\bibinfo {author} {\bibfnamefont {M.}~\bibnamefont
  {Campostrini}}, \bibinfo {author} {\bibfnamefont {P.}~\bibnamefont {Rossi}},\
  and\ \bibinfo {author} {\bibfnamefont {E.}~\bibnamefont {Vicari}},\ }\href
  {https://doi.org/10.1103/PhysRevD.52.358} {\bibfield  {journal} {\bibinfo
  {journal} {Phys. Rev. D}\ }\textbf {\bibinfo {volume} {52}},\ \bibinfo
  {pages} {358} (\bibinfo {year} {1995})}\BibitemShut {NoStop}%
\bibitem [{\citenamefont {Gubser}\ \emph {et~al.}(2019)\citenamefont {Gubser},
  \citenamefont {Jepsen}, \citenamefont {Ji}, \citenamefont {Trundy},\ and\
  \citenamefont {Yarom}}]{Gubser:2019uyf}%
  \BibitemOpen
  \bibfield  {author} {\bibinfo {author} {\bibfnamefont {S.~S.}\ \bibnamefont
  {Gubser}}, \bibinfo {author} {\bibfnamefont {C.~B.}\ \bibnamefont {Jepsen}},
  \bibinfo {author} {\bibfnamefont {Z.}~\bibnamefont {Ji}}, \bibinfo {author}
  {\bibfnamefont {B.}~\bibnamefont {Trundy}},\ and\ \bibinfo {author}
  {\bibfnamefont {A.}~\bibnamefont {Yarom}},\ }\href
  {https://doi.org/10.1007/JHEP09(2019)005} {\bibfield  {journal} {\bibinfo
  {journal} {J. High Energy Phys.}\ }\textbf {\bibinfo {volume} {09}},\
  \bibinfo {pages} {005}}\BibitemShut {NoStop}%
\bibitem [{\citenamefont {Hawashin}\ \emph {et~al.}(2025)\citenamefont
  {Hawashin}, \citenamefont {Eichhorn}, \citenamefont {Janssen}, \citenamefont
  {Scherer},\ and\ \citenamefont {Ray}}]{RayNatCommun}%
  \BibitemOpen
  \bibfield  {author} {\bibinfo {author} {\bibfnamefont {B.}~\bibnamefont
  {Hawashin}}, \bibinfo {author} {\bibfnamefont {A.}~\bibnamefont {Eichhorn}},
  \bibinfo {author} {\bibfnamefont {L.}~\bibnamefont {Janssen}}, \bibinfo
  {author} {\bibfnamefont {M.~M.}\ \bibnamefont {Scherer}},\ and\ \bibinfo
  {author} {\bibfnamefont {S.}~\bibnamefont {Ray}},\ }\href
  {https://doi.org/10.1038/s41467-024-54884-w} {\bibfield  {journal} {\bibinfo
  {journal} {Nat. Commun.}\ }\textbf {\bibinfo {volume} {16}},\ \bibinfo
  {pages} {20} (\bibinfo {year} {2025})}\BibitemShut {NoStop}%
\end{thebibliography}
%% paste by hand from .bbl for submission to Phys Rev

\end{document}

\title{%
Supplemental Material:\\
Non-Lagrangian phases of matter from Wilsonian renormalization of 3D Wess-Zumino-Witten theory on Stiefel manifolds
}

\author{Shouryya Ray}
\email{shouryyar@setur.fo}
\affiliation{Department of Science and Technology, University of the Faroe Islands, Vestara Bryggja 15, 100 T\'orshavn, Faroe Islands}
%%%%%%%%%%%%%%%%%%%%%%%%%%%%%%%%%%%%%%%%%%%%%%%%%%%%%%%%%%%%%%%%%%%%%%%

\date{\today}

\maketitle

%%%%%%%%%%%%%%%%%%%%%%%%%%%%%%%%%%%%%%%%%%%%%%%%%%%%%%%%%%%%%%%%%%%%%%%
\begin{widetext}
\section{Derivation of beta functions}
\subsection{NLSM sector from Ricci flow}
At one-loop, for a general NLSM for $\Phi \colon \mathbb{R}^D \to M$ given by $S = \frac12 \int_x h(\Phi)(\partial_\mu \Phi,\partial_\mu \Phi)$, the one-loop flow equation is most economically derived by exploiting its relation to Ricci flow,
\begin{align}
    \dot{h} = -\dot{\hat{G}}_0(0;\kappa)\operatorname{Ric}_h.
\end{align}
Originally derived for $D=2+\epsilon$ \cite{friedan80,friedan85}, the expression is valid for general $D$ \cite{CodelloPercacci}, where the specific form of the proportionality factor comes from expressing the ``$Q$-functions'' (cf. \emph{ibid.}) in position space. Consider now the expression for the NLSM with Stiefel-manifold target space,
\begin{align}
    S_{\text{NLSM}} = \frac{1}{2g}\int_{\mathbb{R}^D} \operatorname{tr}\!\left(\partial_\mu n^\top \partial^\mu n - \frac{2\alpha + 1}{2\alpha + 2} \partial_\mu n^\top nn^\top \partial^\mu n\right)d^Dx
\end{align}
Take now a tanget vector $\varphi \in T_{I_{N,N-D-1}}\operatorname{St}_{N,N-D-1}$ and write $n = \operatorname{Exp}_{I_{N,N-D-1}}(\varphi)$. Truncating to quadratic order in $\varphi$ yields
\begin{align}
    S_{\text{NLSM}} = \frac{1}{2g}\int_{\mathbb{R}^D} \operatorname{tr} \left(\partial_\mu\phi^\top \partial^\mu\phi + \frac{1}{2(1 + \alpha)} \partial_\mu A^\top \partial^\mu A\right)d^Dx
\end{align}
with the decomposition $\varphi = \begin{pmatrix}A \\ \phi\end{pmatrix}$ where $A \in \mathfrak{so}(N-D-1)$ (in its standard representation) and $\phi \in \mathbb{R}^{(D-1)\times (N-D-1)}$. This allows one to read off for $\xi = \begin{pmatrix}A_\xi \\ \phi_\xi \end{pmatrix}$
\begin{align}
    h(\xi,\xi) = \frac{1}{g}\operatorname{tr} \left(\phi_\xi^\top \phi_\xi + \frac{1}{2(1 + \alpha)} A_\xi^\top A_\xi\right).
\end{align}
The corresponding Ricci curvature is given by \cite{Nguyen}
\begin{align}
    \operatorname{Ric}_h(\xi,\xi) &= \frac{1}{g}\operatorname{tr} \left\{\left[(N-2) + \frac{(N-D-2)}{2(1+\alpha)} \right]\phi_\xi^\top \phi_\xi \right. \nonumber\\
    &\hphantom{{}={}} \left. {} + \left[\frac{N-D-3}{4}\frac{D+1}{4(1 + \alpha)^2}\right] A_\xi^\top A_\xi\right\}.
\end{align}
Comparing coefficients and inserting the expression for $\hat{G}_0(x;\kappa)$ yields the flow equations (at vanishing WZW level, $k=0$) for $g$ and $\alpha$ asserted in the main text.

\subsection{Contribution of the WZW term}
The lowest-order vertex coming from expanding $S_\text{WZW}$ is a $\phi^D$ vertex. To evaluate it, fix $\hat{n}(x,u) = \operatorname{Exp}_{I_{N,N-D-1}}(u \varphi(x))$ to obtain
\begin{align}
    S_\text{WZW} = \frac{2\pi i k}{(D+1)! \Omega_{D+1}}\int_{\mathbb{R}^D} \epsilon^{c_1 \ldots c_{D+1}} \epsilon^{\mu_1 \ldots \mu_D} \delta_{b_{D+1} b_1} \delta_{b_{D-1} b_{D}} \phi_{b_{D+1}}^{c_{D+1}} \partial_{\mu_1}\phi_{b_1}^{c_1} \cdots \partial_{\mu_D}\phi_{b_D}^{c_D}\,d^D x,
\end{align}
where $\varphi = \begin{pmatrix}A \\ \phi\end{pmatrix}$, $\phi = \left(\phi_b^c\right)$ with $b = 1,\ldots,N-D-1$ and $c = 1,\ldots,D+1$. (The field $A \in \mathfrak{so}(N-D-1)$ does not appear at this order.) The contribution of this vertex to the self-energy of the $\phi$ fields is given by the $(D-1)$-loop sunset diagram, which is most efficiently evaluated in position space \cite{Groote:1998wy}. Performing the algebra for the spacetime and internal indices leaves us with
\begin{align}
    \includegraphics[scale=1,valign=c]{Feynman.pdf} = (-1)^{D}p^2 \delta_{bb'}\delta^{cc'} \frac{(2\pi k)^2 g^D}{\Omega_{D+1}^2} \frac{1}{D} \int_x %
    \det\begin{pmatrix} %
    \delta^{\mu_1}_{\mu'_1} & \ldots & \delta^{\mu_1}_{\mu'_{D-1}} \\
    \vdots & \ddots & \vdots \\
    \delta^{\mu_{D-1}}_{\mu'_1} & \ldots & \delta^{\mu_{D-1}}_{\mu'_{D-1}}
    \end{pmatrix} \hat{G}_{0,\mu_1}\hat{G}_{0,}^{\hphantom{0,}\mu'_1} \hat{G}_{0,\mu_{2}}^{\hphantom{0,}\mu_2'} \cdots \hat{G}_{0,\mu_{D-1}}^{\hphantom{0,}\mu'_{D-1}}
\end{align}
It is worth noting that the above expression makes sense for general $D$, though the theory itself is defined only when $D$ is an odd integer. For $D=2$, it leads to
\begin{align}
    \includegraphics[scale=1,valign=c]{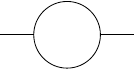} = p^2 \delta_{bb'}\delta^{cc'} k^2 g^2 \frac{1}{2\pi^2} \int_{\mathbb{R}^2} \partial_{\ln \kappa} (\hat{G}_{0,\mu}(x;\kappa))^2 d^2x = p^2 \delta_{bb'}\delta^{cc'} k^2 g^2 \frac{1}{2\pi^2} \int_{\mathbb{R}^2} \partial_{\ln \kappa} l^2 G_0(l;\kappa)^2 \frac{d^2 l}{(2\pi)^2}
\end{align}
Defining now the function $\tilde{r} \colon \mathbb{R}_+ \to \mathbb{R}_+$ by $G_0(l;\kappa)^{-1} \eqqcolon [1 + \tilde{r}(l^2/\kappa^2)]l^2$, the integrand becomes
\begin{align}
    \partial_{\ln \kappa} l^2 G_0(l;\kappa)^2 = \frac{4}{\kappa^2}\frac{\tilde{r}'(l^2/\kappa^2)}{[1+\tilde{r}(l^2/\kappa^2)]^3}.
\end{align}
Making the variable substitution $\tilde{r}(l^2/k^2) = \ell$, whence $d\ell d\Omega_2 = 2 \tilde{r}'(l^2/\kappa^2)/\kappa^2 |l| d|l| d\Omega_2 = 2 \tilde{r}'(l^2/\kappa^2)/\kappa^2 d^2l$, shows $C_{2r}$ to be independent of $\tilde{r}$, and thereby of $r$. A similar trick of Fourier transforming to momentum space, and using that
\begin{align}
    \dot{\hat{G}}_0(0;\kappa) = \int_{\mathbb{R}^2} \dot{G}_0(l^2;\kappa) \frac{d^2l}{(2\pi)^2} = -\frac{2}{\kappa^2}\int \frac{\tilde{r}'(l^2/\kappa^2)}{[1+\tilde{r}(l^2/\kappa^2)]^2} \frac{d^2 l}{(2\pi)^2}
\end{align}
shows that $C_1$ is independent of $\tilde{r}$, and thereby of $r$, in $D=2$. (Note that the position-space representation that is more useful in $D>2$ becomes awkward in $D=2$ due to poles in factors of the form $\Gamma((D-2)/2)$, which cancel subtly against zeros arising from factors of $(x^2)^{(D-2)/2}$.) %
This leads to the beta function
\begin{align}
    \beta_g = -\left[(N-2) - (N-4)\alpha\right]\frac{g^2}{2\pi} + \frac{k^2 g^4}{4\pi^3}
\end{align}
For $N=4$, the level-$k$ $\operatorname{SO}(4) \cong (\operatorname{SU}(2) \times \operatorname{SU}(2))/\mathbb{Z}_2$ WZW theory is recovered, with the one-loop beta function
\begin{align}
    \beta_g = -\frac{g^2}{\pi} + \frac{k^2 g^4}{4\pi^3} = -\frac{g^2}{\pi}\left(1 - \frac{g^2}{4\pi^2/k^2}\right);
\end{align}
the value of the IR-attractive fixed point, $g_* = 2\pi/|k|$, in fact obtains no corrections beyond one-loop, according to a celebrated result of Witten \cite{Witten:1983ar}.
\end{widetext}
%%%%%%%%%%%%%%%%%%%%%%%%%%%%%%%%%%%%%%%%%%%%%%%%%%%%%%%%%%%%%%%%%%%%%%%
% 
% \bibliographystyle{apsrev4-2}
% \bibliography{Stiefelsupp}
%%%%%%%%%%%%%%%%%%%%%%%%%%%%%%%%%%%%%%%%%%%%%%%%%%%%%%%%%%%%%%%%%%%%%%%
%apsrev4-2.bst 2015-08-30 from 4.21a (PWD, AO, DPC/HNN) hacked